# Influence of temporal aspects and age-correlations on the process of opinion formation based on Polish contact survey

Andrzej Grabowski
Central Institute for Labour Protection - National Research Institute
Warsaw, Poland
Email: angra@ciop.pl

Andrzej Jarynowski
Smoluchowski Institute, Jagiellonian University
Cracow, Poland
Email: andrzej.jarynowski@sociology.su.se

*Abstract*—On the basis of the experimental data concerning interactions between humans the process of Ising-based model of opinion formation in a social network was investigated. In the paper the data concerning human social activity, i.e. frequency and duration time of interpersonal interactions as well as age correlations - homophily are presented in comparison to base line homogeneous, static and uniform mixing. It is known from previous studies that number of contact and average age of nearest neighbors are highly correlated with age of an individual. Such real, assortative patterns usually speed up processes (like epidemic spread) on the networks, but here it only plays a role for small social temperature values (by reducing 'freezing by heating' effect). A real structure of contacts affects processes in many various studies in different way, however here it causes stronger (dynamic) and smoother (durations) susceptibility on external field. Moreover, our research shows that the cross interactions between contact frequency and its duration impose the significant increase in critical temperature.

## I. Introduction

Studying the statistical properties of real-world social networks (not only digitized fingerprints of human activities), e.g. contact networks, remains a challenge. In recent years the investigations of social phenomena have been increasingly attracting the interest of the physics community [1], [2], [3]. To assess the basic properties of a network, e.g. the form of the degree distribution, a survey has to be conducted. Using egocentric network dataset (which unfortunately contain only peoples closest neighbors), we collect explicit connections (not implicit as digital) to understand how people manage their personal contact in real life. However, there is still an unexplored area of research concerning human dynamics [4], [5], [6], [7]. Human dynamics has a strong influence on the processes of spreading in social networks [8], e.g. the intercontact times with heavy-tailed distribution affect the spreading of virals [9]. Strong heterogeneity and mixing patterns are some of the most important properties of a social network [10], [11], [12], because they strongly influence the dynamic phenomena in social networks. Assortative mixing the tendency of individuals to associate and bond with similar people is very often observed in many types of social networks. This phenomenon is also known as homophily and the characteristics of agents, such as age, have such a property in our dataset. The mean age of a contact is almost proportional to the age of a participant and most contacts take place within an equal age [13]. Contact heterogeneity should be included in a realistic modelling of opinion formations. Moreover, individuals in real life interact with each other in a different time-scale and in different ways. Note that the model described below allows us to generate a network, of which the structure is based on experimental data, and to consider the real distribution of the duration and frequency of contacts, as well as the real age-age clustering. The two main new features of a social network in our model. Temporal patterns exist in contact networks, because there are people we meet on a daily/weakly/yearly basis and other people we meet randomly. The recent research showed that the real distribution of time between contacts could both slow down or accelerate the spreading of the content, so it is hard to generalise on the relationship between the temporal process and the model behind it [14]. We observe many phenomena related to the temporal aspects, such as the burstiness [15] or fidelity [16]. Recently the availability of temporal contacts (mostly digitalized) is providing critical information to understanding propagation of information. To extract knowledge from this wealth of data an interdisciplinary approach is necessary, combining sophisticate computational modeling with network theory approaches [17]. Motivated by this theoretical studies of temporal networks, we addressed those issues in the model of opinion formation.

Several models are used to investigate various aspects of opinion formation, e.g. the process of the agents adopting new opinions or ideas [18], or to study the effect of heterogeneity on the behaviour of a network [19]. There are different models of opinion formation, the Ising-based model of opinion formation [20] and the Sznajd model adaptation of that model [21] are some of the best known models, especially in the physicist community. In both models, the state of an individual takes a discrete value; however, in some cases, continuous opinion formation models are more useful. The existing mathematical models are sometimes extended to enable the investigation of the influence of an individual's authority or different access to information.

The basic focus of this paper is the process of opinion formation in real social networks [22] through:
1) investigating the interactions between the duration and frequency of contacts (the temporal model);
2) analysing the influence of the age A of each individual preference (homophily) on the structure of a social network (the age-correlated model).

The previous studies of the epidemic model showed that within the SIR model, age-correlations, as well as dynamic and weighed adjustments, increase the total size of outbreaks [23], [24], but the impact on opinion formation is still unknown. To accomplish the mentioned task, numerical simulations were performed using the often applied toy-model: the Ising-based model of opinion formation. The results presented in this paper show the influence of human behaviour, which is the source of the correlations observed in the real world, on the above-mentioned dynamic phenomena. Using real data enables a qualitative and, more importantly, a quantitative assessment of the results. In this paper we present data collected in Poland (1012 participants recorded the characteristics of 16501 contacts with various individuals).

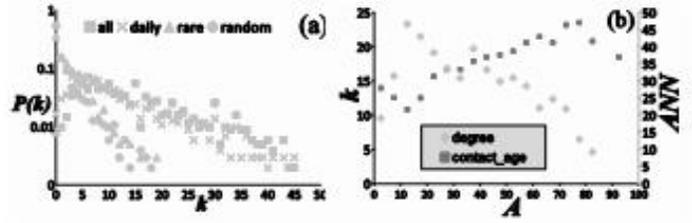

Fig. 2. Empirical degree distribution: all contacts (squares), daily contacts (crosses), rare contacts (triangles), and random contacts (circles) (a), the relationship between the age and the degree and the relationship between the age and the average age of the nearest neighbours (b).

the contacts and the usual frequency of the contacts with the given person. A detailed description of the study is provided elsewhere [22].

On the basis of experimental data [23], the empirical degree distribution of the social network was calculated (Fig.2.a). Initially, the degree distribution increased and had the maximum value of $k = 5$. However, for a large enough $k$ ($k > 5$), the degree distribution had an exponential form $P(k) \sim e^{-\alpha k}$. The value of the exponent $\alpha$ equals 0.07.

The temporal contact patterns under investigation had a three-level structure of interpersonal interactions (Fig.2.a, Fig.I) : (1) all contacts; (2) daily contacts (people we meet almost every day; 72.4% of all contacts); (3) rare contacts (people we meet a few times a month or less frequently; 16.3% of all contacts), and (4) random contacts (people we meet for the first time; 11.3% of all contacts). Initially, the total degree distribution increased and had the maximum value of $k = 5$ (squares in Fig.2.a).

The age A of each individual had a large influenced on the structure of the social network. The connectivity of an individual depended on their age (see Fig.2.b). Maximum connectivity was observed for teenagers and decreased (approximately linearly) with the age increasing. The average age of the nearest neighbours $A_{NN}$ was highly correlated with the age of an individual (see Fig.2.b. For an age greater than 20 years, the $A_{NN}$ increased approximately linearly with the age of an individual increasing. It should be noted that we found similar results in on-line social networks [25]. The demographic structure of the population sampled in the survey corresponds to the Polish Central Statistical Office.

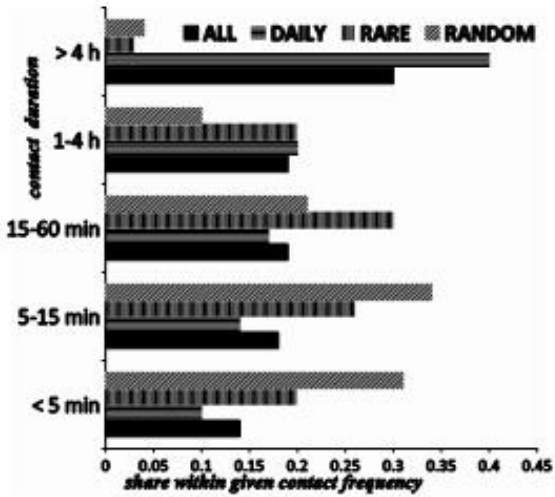

Fig. 1. Weight distributions for various types (intensity) of contacts. Each element of the chart contains the percentage of contacts of a specific type (frequency) and weight (duration), e.g. 14% of all contacts are shorter than 5 minutes.

## II. Data

To construct the model social network, we incorporated data from the contact survey conducted in Poland in the framework of POLYMOD [22]. Quota sampling was applied, taking into account the demographic variables, such as age, sex, region and type of residence. The participants were recruited by trained interviewers, visiting random households and each participant was randomly assigned one day, on which they had to record all their contacts. A contact was counted only in physical presence and the participants had to record all contact episodes with each contact person during the assigned day. Additionally, they filled the total duration, the location of

## III. Network model

The following generic procedure of creating a network was used. For each individual we drew $k_i$ (the desired number of connections from experimental distribution (see Fig.2.a). Note that for different network models we use a different experimental distribution (for more details see Table I and Fig.3). To investigate the influence of contact frequency, we made computations for two types of networks: static and dynamic. Additionally, link weight was taken into account, so that our network could be weighed or uniform. To investigate the influence of age-age and age-degree correlations, we distinguished another dimension: correlated and non-correlated.

TABLE I
THE COMPARISON OF DIFFERENT NETWORK MODELS WITH
CHARACTERISTIC PROPERTIES: DUrations, COrrelations,
CLustering procedure and DYnamics

| Network model | DUrations and COrrelations | CLustering procedure and DYnamics |
|---|---|---|
| static and uniform and non-correlated | both DU and CO uniform | CL all connections and DY fully static |
| static and weighted and non-correlated | DU weighted and CO uniform | CL all connections and DY fully static; |
| dynamic and uniform and non-correlated | both DU and CO uniform | CL daily only, DY rare and random connections re-created at each step |
| dynamic and weighted and non-correlated | DU weighted and CO uniform | CL daily only, DY rare and random connections re-created at each step |
| static and uniform but correlated | DU uniform and CO according to age-age and age-degree empirical distribution | CL all connections with control of age/degree statistics and DY fully static |

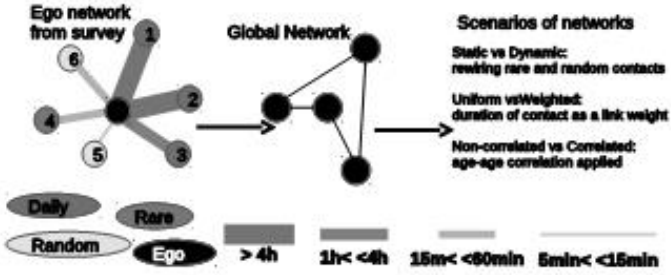

Fig. 3. Construction of the network from ego-networks with various scenarios

All networks included $10^6$ individuals, which was the order of magnitude of a country. At the beginning, for all nodes the actual number of connections was zero ($c_i = 0$). Next, the connections between individuals were created according to different scenarios (Table I).

### A. Temporal model

In a static network, all contacts are created at the beginning of a simulation according to $P_{ALL}(k)$ distribution (squares in Fig.2.a). In a dynamic network, daily contacts are created at the beginning of a simulation only; rare and random contacts are re-created in each time step [26].

The general procedure presented below allowed us to create a network with the properties typical of social networks, e.g. small values of the average shortest paths and large clustering, and it was introduced in our previous paper in the static and the dynamic case [23].

1. A pair of nodes (i and j) is chosen at random from a group of nodes, in which the actual number of connections is less than the desired number of connections, i.e. when $c_i < k_i$. 2. A connection between the pair of nodes is created. If the nodes have been connected previously, go to step 1, because only single connections are allowed.
3. Once a connection has been created (if the network is weighted, its weight is generated with proper distribution $P(w)$ according to Fig. I) - connections are symmetrical, i.e. $w_{ij} = w_{ji}$.
4. In this step, we increase the value of the clustering coefficient $C$. If possible, a new connection between the ith individual and an individual chosen from the neighbours of the neighbours of the ith individual is created - this procedure increases the value of the clustering coefficient $C$. Otherwise, a new neighbour is chosen from the entire population.

The above-mentioned procedure was repeated until $c_i < k_i$ and finally the desirable distribution of connectivity was produced. Here, the clustering coefficient is $C \approx 0.4$ and the average shortest path equals $\langle l \rangle \approx 5.6$. The obtained network is assortatively mixed by degree and hierarchical, without imposing this directly. The individuals with small $k$ run out of all connections in first steps of the simulation, thus the individuals with larger desired $k$ must connect with nodes with similar attributes (highly interconnected because of the high clustering coefficient). Moreover, small groups are hierarchically organised in increasingly large groups.

The procedure of re-wiring [27] rare and random contacts in a dynamic network is as follows. The number of daily contacts for each individual is sampled from $P_{daily}(k)$ (crosses in Fig.2.a) distribution. Note that in a dynamic network, a large proportion of the connections (daily contacts constitute 72.4% of all contacts) does not change during the simulation. At each time step of the simulation, all rare and random contacts are removed and we search for a possible replacement. Re-wired links are following $P_{rare}(k)$ (triangles in Fig.2.a), and $P_{random}(k)$ (circles in Fig.2.a) empirical distributions, respectively.

Rare contacts are sampled from the local network of an individual. So, we preferentially look for neighbours of neighbours in accordance with the clustering procedure. This causes repetitions of rare contacts. Random contacts are always chosen from the entire population. The weight of a new connection is chosen as in step 3 of procedure.

### B. Age-correlated model

The second concept of network adjustment are age-age and age-degree correlations and we developed simple network models using such information. We create individuals according to the demographic age $A$ structure (the individuals are binned in 5-year age groups). The ith individual of the given age $A$ initially has $k_i$ free connections with a desired age of a neighbour (sampled from distribution Fig.2.b).

The connections between individuals are created in a modified procedure from the previous subsection and we add homophily-driven re-wiring. Here too, the individual i can be connected only when its actual number of connections is smaller than the value $k_i$.

1. A pair of nodes (i,j) can be connected only if the ith individual has a free connection to the individual with the of age $A_j$, and the jth individual has free connection to the individual with the age of $A_i$ (because we take into account the age of individuals). A connection between individuals i and j is established.

2. In the next step additionally 'clustering' procedure is applied. We search for a possible new connection between the ith individual and the available neighbours of the jth individual (i.e. having at least one free connection with the proper age assigned).
3. Iterate the previous points to reach the desired distribution of connectivity (the actual number of connections $c_i$ of a node i should be comparable to $k_i$ sampled from the empirical distribution).

Taking into account age-age and degree-age correlations, we create a network with properties, e.g. small average shortest path, medium clustering (0.2), assortative mixing [28], [12].

## IV. ISING-BASED MODEL OF OPINION FORMATION

We study the evolution of opinions (opinion formation). We introduce the local field $h_i$ for each individual, which is a function of interactions with $k_i$ neighbours and the external field (stimulation) I:

$$h_i(t) = -S_i(t)\left(\sum_{j}^{k_i} w_{ij} S_j(t) + I\right) \quad (1)$$

where $S_i = \pm 1$ - state of i-th individual, $k_i$ - number of neighbors of i-th individual (in the case of dynamic network this value change in time), $w_{ij}$ - weight of interaction between i-th and j-th individuals.
The external field I replaces the interaction with all other individuals and may be considered as the influence of mass media.

The opinions of the individuals change simultaneously, according to the probability corresponding to the local field and temperature T:

$$S_i(t+1) = \begin{cases} S_i(t): & \frac{\exp(-h_i/T)}{\exp(-h_i/T)+\exp(h_i/T)} = \frac{1}{1+\exp(2h_i/T)} \\ -S_i(t): & \frac{\exp(h_i/T)}{\exp(-h_i/T)+\exp(h_i/T)} = \frac{1}{1+\exp(-2h_i/T)} \end{cases} \quad (2)$$

The parameter T may be interpreted as the **social temperature**. It describes the degree of randomness of the behaviour of individuals. If the temperature T increases, the probability that the individual will have a state opposite to the local field is higher. Note that Eq.(2) is analogous to Glauber dynamics [29]. The computations were performed for the initial conditions corresponding to the paramagnetic phase ($\langle S \rangle = 0$ for $t = 0$) using synchronous dynamics, while analysing the interdependence between time t and temperature T. We would like to find the critical value of temperature $T_c$ and the time evolution of $\langle S \rangle$ in various settings. When external stimulation I is investigated, initially all individuals states were opposite to external stimulation. The size of the network is always: $10^6$ individuals.

## V. RESULTS: AGE - CORRELATED (HOMOPHILY) VS NON-CORRELATED

We investigate how the correlations between the age of the participants influence the process of opinion formation.

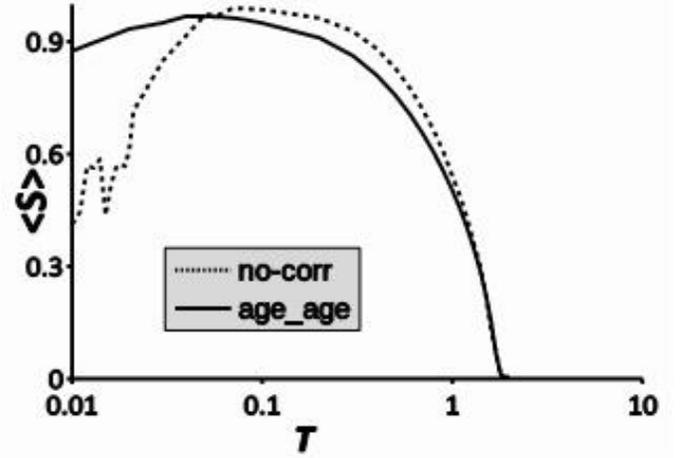

Fig. 4. The relationship between $\langle S \rangle$ and the temperature T for different types of networks (with age-age correlations - solid line and without age-age correlations - dashed line). Simulation time is 1000 steps and results were averaged over $10^2$ independent simulations.

In the relation between $\langle S \rangle$ and temperature T, we observe 3 regimes, below, around and above the critical $T_c$ (Fig.4). The trajectories of both scenarios (correlated homophily) vs the non-correlated do not vary at large temperatures. Due to the state transition formulation (2) if T goes to infinity, the opinion changes randomly and the average opinion is neutral $\langle S \rangle \approx 0$. At low temperatures, the dominant opinion emerges in the community in all cases. When the temperature exceeds a certain critical value $T_C$ there is an abrupt disappearance of the dominant opinion in the community - a phase transition is observed [30]. As we see in Fig. 6, the dominant opinion emerges after approx. 100 time steps, even at temperatures close to the critical value. For all networks, the value of $T_C$ is the same ($T_C \approx 1.85$). Discrepancies in results are observed at lower temperatures, where homophily can both inhibit and catalyze process.

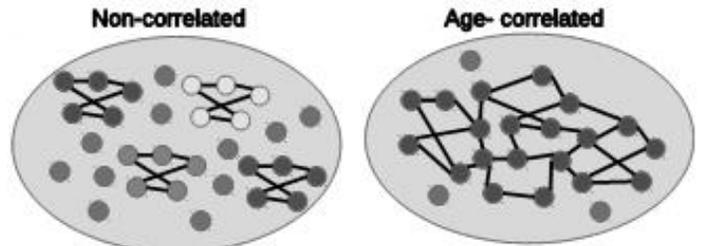

Fig. 5. The difference in the network structure causing non-monotonic relation of $\langle S \rangle$

In the case of network with age-age correlations and the value $T > 0.05$ of $\langle S \rangle$ is smaller. Lower values of magnetisation in the case of a network with age-age correlations may be a result of the presence of groups of individuals that are weekly connected with the rest of the individuals [31]. An example of such group are the elderly, who have a lower number of connections in average, and the most of their connections are in the same age groups. They are weekly connected with age groups with the largest number of connections (see Fig.2.b). Such groups may have the opinion opposite to the opinion of almost the entire population. Hence the value of $\langle S \rangle$ is smaller.

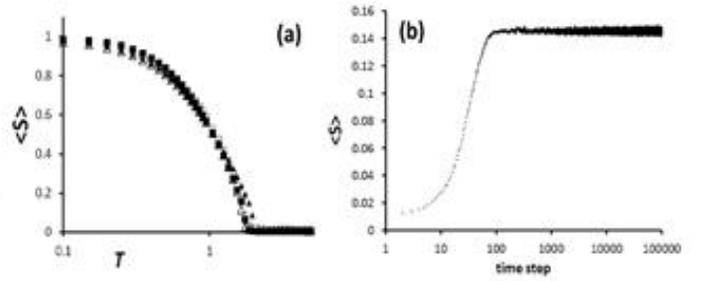

Fig. 7. The relationship between $\langle S \rangle$ and the temperature T for different types of networks (static - white markers, dynamic - black markers) and different distributions of weights (real - triangles, uniform - squares). The relationship between $\langle S \rangle$ and the time for a dynamic network with a real distribution of weights (T = 1.85). The results were averaged over $10^3$ independent simulations.

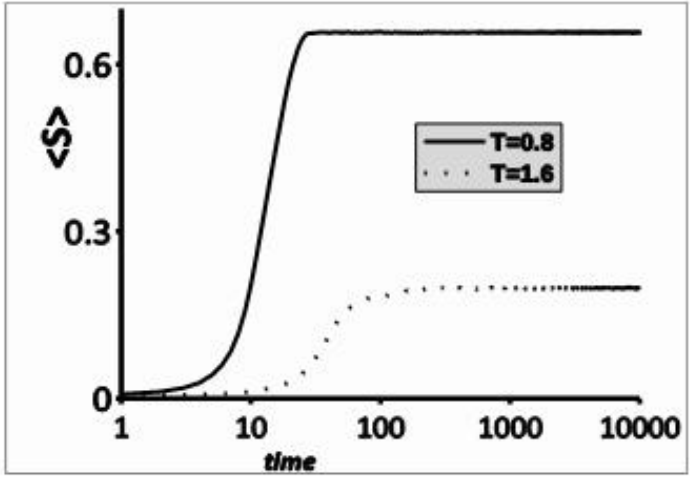

Fig. 6. The relationship between $\langle S \rangle$ and time for different temperatures T (0.8 and 1.6 from top to down, respectively) for network without age-age correlations. Results were averaged over $10^2$ independent simulations.

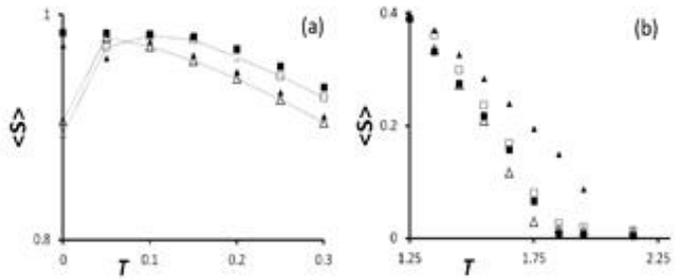

Fig. 8. The relationship between $\langle S \rangle$ and the temperature T for different types of networks (static - white markers, dynamic - black markers) and different distributions of weights (real - triangles, uniform - squares) and for low (a) and high (b) temperatures. The results were averaged over $10^3$ independent simulations.

We observe an interesting behaviour for a T slightly greater than zero. It is visible (Fig.4) that $\langle S \rangle$ initially increases with the temperature increasing. This behaviour is similar to the so-called "freezing by heating" phenomenon (in which the system reaches a ferromagnetic state for a high enough temperature). In the non-correlated networks (Fig. 5), many small communities are formed (the clustering coefficient of the social networks is high). The individuals in such a community are highly interconnected and the number of connections outside the community is relatively small. Therefore it is possible that these communities have a minority opinion (hence $\langle S \rangle$ is smaller). An increase of the temperature causes the opinion of such communities to switch to the dominant opinion - the system can jump from the local to the global minimum of energy. Note that the "freezing by heating" phenomenon is observed in networks with age-age correlations with the lower extent ( $\langle S \rangle$ for T ≈ 0.01 is larger). This is because in such case the dominant component of the network is observed, individuals with the age of A ≈ 20 and a very large number of connections. Those individuals not only quickly chose a common opinion, but they are also able to convince others to choose their opinion. Hence the value of $\langle S \rangle$ is greater for a correlated network.

VI. RESULTS: DYNAMIC (TEMPORAL) VS STATIC

Here, we investigate the effect of the temporal re-wiring of the network. Outside the critical temperature $T_c$ and time evolution, we study the role of the mass media in the process of opinion formation, modelled as external stimulation I acting on the social network.

The relation between $\langle S \rangle$ and the temperature T is shown in Fig.7.a, 8. Differences between networks emerge in low temperatures (T ≈ 0) and temperatures slightly lower than the critical value (T ≈ $T_C$) (see Fig.8).

The differences between networks emerge at low temperatures (T ≈ 0) and at temperatures slightly lower than the critical value (T ≈ $T_C$) (see Fig.8).

For T = 0, it is visible (Fig.8.a) that for static networks (dashed lines), the value of $\langle S \rangle$ is lower and initially increases with the temperature increasing. The presence of the groups of nodes that are highly interconnected by connections with a high value of weights causes an increase in the critical value of the temperature (Fig. 7.b). It turns out that the interactions of temporal aspects: the duration of the contact (the weight of the contacts) with its frequency shifting critical temperature, while one factor alone (duration and frequency) does not change it. The value of an average spin near the critical value of temperature is stable and does not change significantly for $10^5$ time steps. Note that one time step corresponds to one day, hence the state of the network is stable for over three hundred years. We do not observe the "freezing by heating" phenomenon in a dynamic network.

In this paper, the possibility of change in the states of individuals due to external stimulation I is investigated, i.e. we assume that initially all individuals have states opposite to external stimulation. It can be noticed that for I exceeding the critical value of $I_C$, most individuals have the state of $S = +1$ (Fig.9). This means that a certain intensity of the influence of the mass-media can change the opinion of the community to a state constrained by the media.

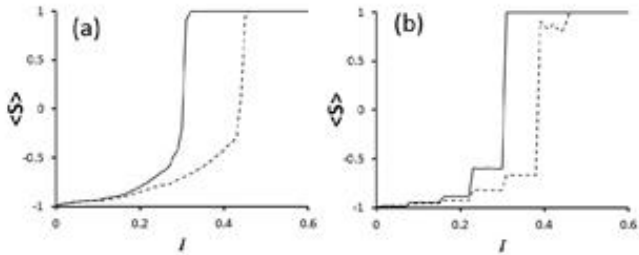

Fig. 9. The relationship between $\langle S \rangle$ and the value of external stimulation I for different types of networks (static - dashed line, dynamic - solid line) and different distributions of weights (real - (a), uniform - (b)). The results were averaged over $10^3$ independent simulations.

For a uniform distribution of weights, the relationship between $\langle S \rangle$ and I has the form of stairs [29]. Because all connections between individuals are the same (because the weights are the same), a network can be divided into smaller groups according to the number of connections. For a high enough value of I, the opinion of all individuals in the group is reverted (first individuals with lowest connectivity $k = 1$ change their opinion, then individuals with $k = 2$, and so on), because the influence of external stimulation is larger than the total influence of the neighbours ($I > k * average weight value$). For a real distribution of weights, the influence of the neighbours is much more diverse, because it depends on the number of connections and on the weights. Hence, it is impossible to distinguish large groups of nodes with the same properties within the network (i.e. the same number of connections and the same number of weights).

Dynamic networks are less resistant to external stimulation - the opinion of the community is changed at lower values of I (cf. solid and dashed lines in Fig.9).

## VII. CONCLUSIONS

Dynamic networks are less resistant to external stimulation - the opinion of the community is changed at lower values of I (cf. solid and dashed lines in Fig.9).

Opinion formation in a social network constructed on the basis of experimental contact data on interactions between humans is studied. In this paper, data on human social activity: temporal aspects of the duration (weight of contacts) with frequency; as well as homophily (age-age correlation) of interpersonal interactions are presented.

By investigating the temperature, various effects are observed at temperatures below and above $T_C$. At $T \approx 0$, initially the average spin increases with the temperature increasing.

This behaviour is similar to the so-called "freezing by heating" phenomenon. In correlated networks this effect is smaller and this is not observed at all in dynamic networks, because a random re-wiring of connections has a similar influence to an increase in temperature (the system is less prone to sticking at a local minimum). In correlated networks the average spin is smaller than in non-correlated networks for $0.05 < T < T_c$. Our research shows that age-age correlations have a small influence on the process of opinion formation mostly at the low temperatures mentioned above. The results of numerical calculations indicate that the age structure does not influence $T_C$, but only the speed of the phase transition. Age-correlations slow down critical processes, so such networks are less prone to abrupt changes [32]. It was showed before [24], that age-correlations in the same survey decrease the total size of an epidemic outbreak, which is not the case of opinion formation. When both temporal aspects (weight and frequency) are considered, the critical value of the temperature $T_C$ increases significantly (Fig.8.b). The reason could be the fact that dynamic links have short durations and could be called weak' ties [33] between the communities that could have a different opinion. Important changes in the behaviour of the system are observed only at low temperatures (slightly greater than 0), at which the "freezing by heating" effect occurs only in static networks and at temperatures close to the critical value $T \approx T_C$.

We test the impact of an external field on our network. For a dynamic network, the critical value of external stimulation necessary to change the opinion of the population is much smaller (Fig.9). Dynamic networks are less resistant to external stimulation because the process of re-wiring introduces randomness to the system. Individuals with the state $S = -1$ can be connected with individuals with the state $S = +1$ and change their opinion as a result. In such systems, the influence of re-wiring is similar to the influence of temperature - the higher the temperature, the more susceptible the network and the lower the value of $I_C$.